\documentclass[final]{aipproc}

\layoutstyle{8x11single}
\newcommand{\lsim}{\raisebox{-0.7ex}{$\stackrel{\textstyle <}{\sim}$ }}
\newcommand{\gsim}{\raisebox{-0.7ex}{$\stackrel{\textstyle >}{\sim}$ }}

\begin{document}

\title{Hadronic Interactions with lattice QCD}


\author{Martin J. Savage~\footnote{for the NPLQCD collaboration}}{
  address={Department of Physics, University of Washington, Seattle, WA 98195-1560.}
}

\begin{abstract}
I will describe recent progress 
toward calculating hadronic interactions with Lattice QCD.
\end{abstract}

\maketitle


\section{Introduction}

It has been known for the past four decades that Quantum
Chromodynamics (QCD), together with the electroweak interactions,
underlies all of nuclear physics.
Soon after the discovery of QCD it became apparent that
the complexities of the theory at strong coupling would hinder
analytic progress in understanding the properties of the simplest
hadrons, let alone the simplest features of the nuclear forces.
Wilson pointed the way to eventual
direct quantitative confirmation of the origins of nuclear physics by
formulating Lattice QCD~\cite{Wilson:1974sk}, a regularization 
and non-perturbative definition
of QCD
that is suitable for the intensive computational demands of solving
QCD in the infrared.
Only in the last five years or so has Lattice QCD emerged from a long
period of research and development ---where only qualitative agreement
between calculations and experiments could be claimed--- to the
present, where precise predictions for hadronic quantities are being
made. In particular, presently, fully-dynamical calculations with
near-exact chiral symmetry at finite lattice-spacing have become standard,
with lattice volumes of spatial extent $L \gsim 2.5~{\rm fm}$ and
with lattice spacings in the range $b\lsim 0.12~{\rm fm}$. 
It is still the norm that the light-quark
masses, $m_q$, are larger than those of nature, with typical pion
masses $m_\pi\sim 300~{\rm MeV}$.

Nuclear physics is a vast, rich field, whose phenomenology has been
explored for decades through intense experimental and theoretical
effort. However, there is still little understanding of the connection
to QCD and the basic building blocks of nature, quarks and gluons.
For instance, as a first ``benchmarking'' step, Lattice QCD should
post-dict the large nucleon-nucleon scattering lengths and 
the existence of the deuteron, the simplest nucleus.  The
connection between QCD and nuclear physics will be firmly established
with Lattice QCD, and will allow for an exploration of how nuclei and
nuclear interactions depend upon the fundamental parameters of
nature. In particular, it is believed that an understanding of the
fine-tunings that permeate nuclear physics will finally be translated
into fine-tunings of the light-quark masses. While these
issues are of great interest, and it is important to recover what is
known experimentally to high precision, these goals are not the main
objective of the Lattice QCD effort in nuclear physics.  The primary
reason for investing resources in this area
is to be able to calculate physical quantities of importance that
cannot be accessed experimentally, or which can be measured with only
limited precision in the laboratory. Two important examples of how
Lattice QCD calculations can impact nuclear physics are in the
structure of nuclei and in the behavior of hadronic matter at 
densities beyond that of nuclear matter.
Lattice QCD will be able to calculate the
interactions of multiple nucleons, bound or unbound, in the same way that it
can be used to determine the two-body scattering parameters.  For
instance, a calculation of the three-neutron interactions will be
possible.

Extracting hadronic interactions from Lattice QCD calculations is
more complicated than the determination of the spectrum of stable particles.
This is encapsulated in the
Maiani-Testa theorem~\cite{Maiani:1990ca}, which states that S-matrix elements
cannot be extracted from infinite-volume Euclidean-space Green functions except
at kinematic thresholds.
This is clearly problematic from the nuclear physics perspective, as a main
motivation for pursuing Lattice QCD is to be able to compute nuclear reactions
involving multiple nucleons.
Of course, it is clear from the statement of this theorem how it can be evaded,
Euclidean-space correlation functions are calculated at finite volume to extract
S-matrix elements, the formulation of which was known for decades in the
context of non-relativistic quantum mechanics~\cite{Huang:1957im} and extended
to 
quantum field theory
by L\"uscher~\cite{Luscher:1986pf,Luscher:1990ux}.
The energy of two particles in a  finite volume depends
in a calculable way upon their elastic scattering amplitude and their masses
for energies below the inelastic threshold. 

\section{Meson-Meson Interactions}

Due to the chiral symmetry of
QCD, $\pi\pi$ scattering at low energies is the simplest and
best-understood of the hadron-hadron scattering processes.  The scattering
lengths for $\pi\pi$ scattering in the s-wave are uniquely predicted
at LO in $\chi$-PT~\cite{Weinberg:1966kf}:
$
m_\pi a_{\pi\pi}^{I=0} \ = \ 0.1588 $ and 
$m_\pi a_{\pi\pi}^{I=2} \ = \
-0.04537 $
when $m_\pi$ is set equal to the charged pion mass. While experiments 
do not directly provide stringent
constraints on the scattering lengths, a determination of s-wave
$\pi\pi$ scattering lengths using the Roy equations has reached a
remarkable level of
precision~\cite{Colangelo:2001df,Leutwyler:2008fi}:
$m_\pi a_{\pi\pi}^{I=0} \ = \ 0.220\pm 0.005$
and
$m_\pi a_{\pi\pi}^{I=2} \ = \ -0.0444\pm 0.0010$.
At present, Lattice QCD has computed $\pi\pi$ scattering
only in the $I=2$ channel with precision 
as the $I=0$ channel contains disconnected
diagrams. 
The only existing $n_f=2+1$ Lattice QCD prediction of the
$\pi^+\pi^+$ scattering length involves a mixed-action Lattice QCD scheme
of domain-wall valence quarks on a rooted staggered sea~\cite{Beane:2007xs}.
The scattering length was computed at pion masses, $m_\pi\sim 290~{\rm
  MeV}$, $350~{\rm MeV}$, $490~{\rm MeV}$ and $590~{\rm MeV}$, and at
a single lattice spacing, $b\sim 0.125~{\rm fm}$ and lattice size
$L\sim 2.5~{\rm fm}$~\cite{Beane:2007xs}. The physical value of the
scattering length was obtained using two-flavor mixed-action $\chi$-PT
which includes the effect of finite lattice-spacing
artifacts to $\mathcal{O}(m_\pi^2 b^2)$ and
$\mathcal{O}(b^4)$~\cite{Chen:2006wf}.  The final result is:
$m_\pi a_{\pi\pi}^{I=2}  =   -0.04330 \pm 0.00042$.  
Lattice QCD calculations at one or more smaller lattice spacings are
required to verify and further refine this calculation.
Recently, the ETM
collaboration has performed a $n_f=2$ calculation of
the $\pi^+\pi^+$ scattering length at
pion masses ranging from $m_\pi\sim 270~{\rm MeV}$ to $485~{\rm MeV}$,
at two lattice spacings  and in two lattice volumes. 
The result extrapolated to 
the physical pion mass is:
$m_\pi a_{\pi\pi}^{I=2}  =   -0.04385 \pm 0.00028 \pm 0.00038$.

\begin{figure}[!ht]
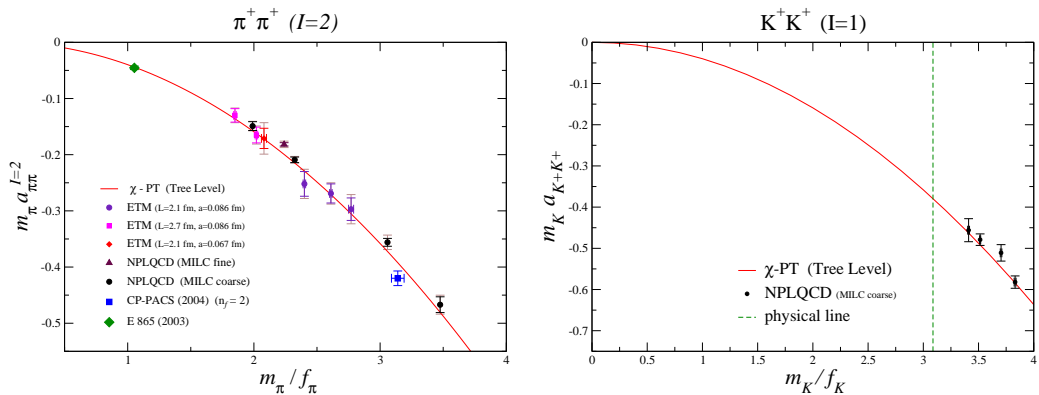

  \includegraphics[height=0.23\textheight]{figures/mpia2Plot4_pipi.eps}\qquad\includegraphics[height=0.23\textheight]{figures/muaCA_PLOT_KK.eps}
  \caption{$m_\pi a_{\pi^+\pi+}$ vs. $m_\pi/f_\pi$ (left panel)
and $m_k a_{K^+K^+}$ vs. $m_k/f_k$ (right panel). The black points
are the results of  Lattice QCD calculations by 
the NPLQCD collaboration, the blue point (left panel) is 
from the CP-PACS collaboration,
the purple, red and magenta points (left panel) 
are from the ETM collaboration.
The
red lines are the current algebra predictions.
}
\label{fig:CAplots}
\end{figure}

In Figure~\ref{fig:CAplots} (left panel) one sees that the
$\pi^+\pi^+$ scattering length is consistent with 
the current algebra result up
to pion masses that are expected to be at the edge of the chiral
regime in the two-flavor sector. 
While in the two flavor theory one
expects fairly good convergence of the chiral expansion and, moreover,
one expects that the effective expansion parameter is small in the
channel with maximal isospin, the lattice calculations clearly imply a
cancellation between chiral logs and counterterms. 
However, as one sees in Figure~\ref{fig:CAplots} (right
panel), the same phenomenon occurs in $K^+K^+$ where the chiral
expansion is governed by the strange quark mass and is therefore
expected to be much more slowly converging.  The $\pi^+K^+$ scattering
length exhibits similar behavior when $\mu_{K\pi}\ a_{K^+\pi^+}$ is
plotted against $\mu_{K\pi}/\sqrt{f_K f_\pi}$.  
This cancellation between
chiral logs and counterterms for the meson-meson scattering lengths is
quite mysterious.
\vskip 0.1in

\section{Baryon-Baryon Interactions}

\begin{figure}[!ht]
  \includegraphics[height=0.22\textheight]{figures/SING_NNlattice2009.eps}\ \ \ \ \ \ \ \includegraphics[height=0.22\textheight]{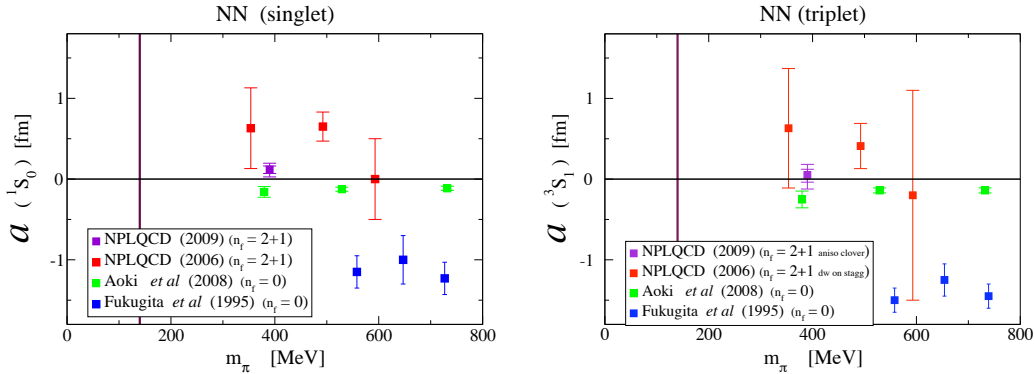}
  \caption{A compilation of the scattering lengths for NN scattering
    in the $^1S_0$ (left panel) and $^3S_1$ (right panel) calculated
    with Lattice QCD and with quenched Lattice QCD.  
The vertical lines correspond to
    the physical pion mass.
}
\label{fig:NN-ALL-LQCD}
\end{figure}

The NPLQCD collaboration performed the first $n_f=2+1$ QCD calculations
of nucleon-nucleon interactions~\cite{Beane:2006mx} and
hyperon-nucleon~\cite{Beane:2006gf} interactions at low-energies but with
unphysical pion masses, and  the nucleon-nucleon scattering lengths were found
to be of natural size. 
Such calculations are far more challenging than those involving mesons due to
the exponential degradation of the signal-to-noise at large times.
A summary of all Lattice QCD calculations of NN scattering is shown in
Figure~\ref{fig:NN-ALL-LQCD}.  
The results calculated 
on the 
anisotropic clover gauge configurations
are consistent with those obtained with
mixed-action Lattice QCD~\cite{Beane:2006mx}, and it is interesting to
note that the results of quenched calculations~\cite{Aoki:2008hh}
yield scattering lengths that are consistent within uncertainties with
the fully-dynamical $n_f=2+1$ values.
\vskip 0.1in

\begin{figure}[!ht]
  \includegraphics[height=0.21\textheight]{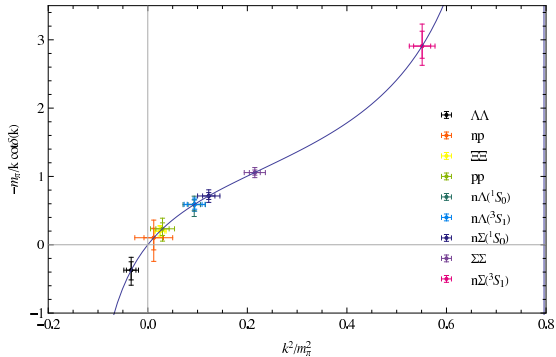}
  \caption{Baryon-baryon interactions extracted from calculations at 
$m_\pi\sim 390~{\rm MeV}$~\cite{Beane:2009py}.
}
\label{fig:BB}
\end{figure}

Figure~\ref{fig:BB} shows a summary of the high-statistics  
results for baryon-baryon
interactions recently
obtained on the $20^3\times 128$ anisotropic clover gauge configurations with
$m_\pi\sim 390~{\rm MeV}$.  The
$\Lambda\Lambda$ channel is found to be negatively shifted in energy which may
signal that the lowest-state is in fact a bound-state, but
calculations in a larger volume are required before more definitive
conclusions can be drawn. 
Calculations at this pion mass should be viewed to be only a
proof of principle, and 
in order for lattice calculations to provide meaningful constraints 
on interactions at the physical pion mass through
chiral extrapolation, the pion mass must be substantially reduced down
toward the physical value ($m_\pi\sim 140~{\rm MeV}$) while
maintaining the integrity of the calculation (i.e. small enough
lattice spacings and large enough volumes).  
\vskip 0.1in

I should also mention that there have been calculations of energy-dependent
and source-dependent ``potentials'' that are constructed in such a way to
reproduce the calculated phase-shifts at the energies of the states in the
finite lattice volumes, for instance, see Ref.~\cite{Inoue:2010hs}.  
Currently unjustified assumptions are required in
order for these objects to be compared with the modern
nucleon-nucleon or hyperon-nucleon potentials that are used in the calculation
of nuclear and hyper-nuclear structure and interactions, and I will discuss
them no further.

\section{Multi-Baryon Interactions}

\begin{figure}[!ht]
  \includegraphics[height=0.22\textheight]{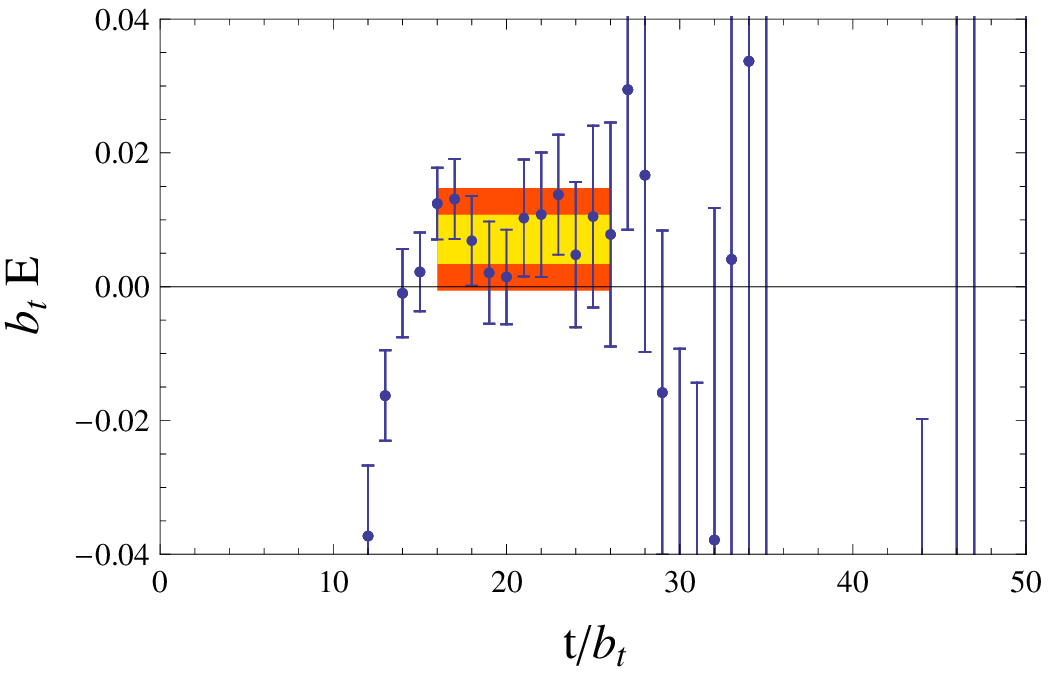}\ \ \ \ \ \ \ \includegraphics[height=0.27\textheight]{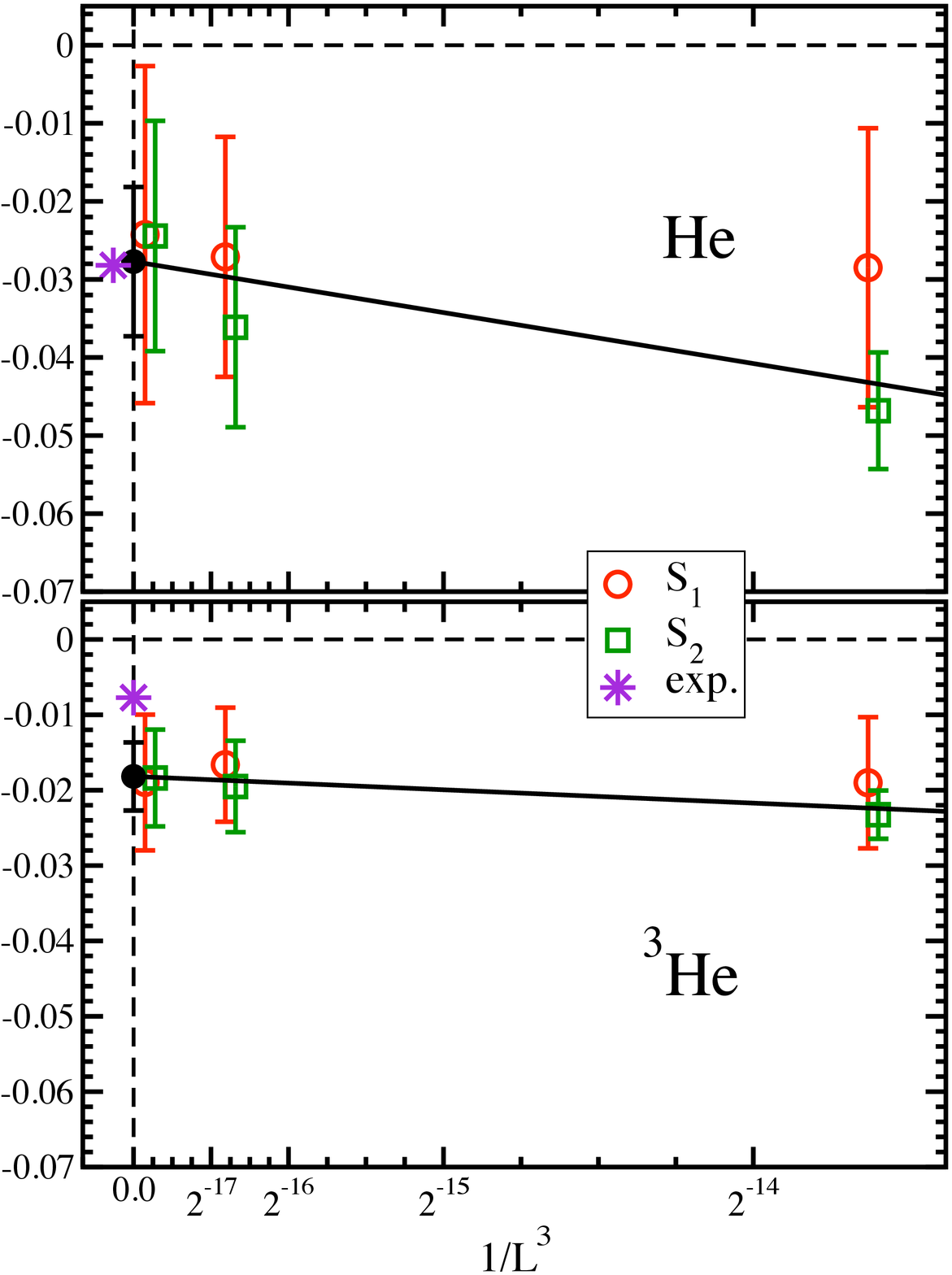}
  \caption{
The left panel shows the 
effective mass plot in the triton channel resulting from
the $n_f=2+1$ calculation with $m_\pi\sim 390~{\rm MeV}$
after the subtraction of $3\times M_N$.
The right panel shows the quenched results 
for the binding energies (in lattice units)
obtained by the PACS-CS collaboration in the 
channels with the quantum numbers of the 
triton and the $\alpha$-particle~\cite{Yamazaki:2009ua} with 
$m_\pi\sim 800~{\rm MeV}$.
}
\label{fig:threeplus}
\end{figure}

The first significant steps towards the calculation of the properties of 
nuclei directly
from QCD were taken by the NPLQCD collaboration~\cite{Beane:2009gs}
and the PACS-CS collaboration~\cite{Yamazaki:2009ua}
during 2009.  
The NPLQCD collaboration performed a $n_f=2+1$ calculation  
of the correlation function with the quantum
numbers of the strangeness $s=-4$, baryon number $B=3$ system, which is labeled
as ``$n\Xi^0\Xi^0$'' for convenience, 
and also of the correlation function with the quantum numbers 
of the triton (or $^3$He),
at a pion mass of $m_\pi\sim 390~{\rm MeV}$~\cite{Beane:2009gs} as shown in 
fig.~\ref{fig:threeplus}. 
Further, the PACS-CS collaboration performed a quenched calculation of 
the system with the quantum numbers of the triton and of the 
$\alpha$-particle at a pion mass of 
$m_\pi\sim 800~{\rm MeV}$~\cite{Yamazaki:2009ua}  as shown in 
fig.~\ref{fig:threeplus}. 

The dream of being able to perform reliable
calculations of the interactions among  multiple nucleons and 
hyperons, and of the
structure and reactions of light-nuclei,  directly from QCD
is starting to be realized. 
The path forward is clear, and the next decade will be a truly remarkable 
period for nuclear physics.

\begin{theacknowledgments}
I would like to thank all of the members of the NPLQCD collaboration 
for their contributions to this program.
\end{theacknowledgments}

\end{document}